\documentclass[%reprint,
superscriptaddress,
preprint,
 amsmath,amssymb,
 aps,
]{revtex4-1}

\usepackage{graphicx}% Include figure files
\usepackage{dcolumn}% Align table columns on decimal point
\usepackage{hyperref}
\usepackage{bm}% bold math
\usepackage{enumerate}
\usepackage{lipsum}
\usepackage{epstopdf}
\usepackage{float}
\usepackage[caption = false]{subfig}
\usepackage{tabularx}
\usepackage{mathtools}

\begin{document}

\preprint{APS/123-QED}

\title{Interaction of a Laguerre-Gaussian beam with Rydberg atoms}

\author{Koushik Mukherjee}
\author{Sonjoy Majumder}
\email{sonjoym@phy.iitkgp.ernet.in}
\affiliation{Department of Physics, Indian Institute of Technology Kharagpur, Kharagpur-721302, India.}
\author{Pradip Kumar Mondal}
\affiliation{Department of Physics, Supreme Knowledge Foundation Group of Institutions, Hooghly - 712139, India.}
\author{Bimalendu Deb}
\affiliation{Department of Materials Science, Indian Association for
the Cultivation of Science, Jadavpur, Kolkata 700032, India}

\date{\today}


%\affiliation{
%\affiliation{
 %Third institution, the second for Charlie Author
%\affiliation{%



\begin{abstract}

Transfer mechanism of orbital angular moment(OAM) of light to trapped ground-state atoms under paraxial approximation is well known. Here we show how optical OAM of a Laguerre-Gaussian(LG) beam under paraxial approximation can be transferred to trapped Rydberg atoms. Optical OAM is shown to be transferable to a Rydberg electronic state in dipole transition. The Gaussian part of the profile of the LG beam,  which is generally neglected , is found to have an important effect on the OAM transfer to the Rydberg atoms. Numerical calculations are calculated based on this theory for Rubidium Rydberg atoms trapped in a harmonic potential. Our results exhibit the mixing of final states of different parities.

\end{abstract}

\maketitle

\section{INTRODUCTION}

It has been twenty five years since the seminal work by Allen et al.\cite{Allen1992} showing that Laguerre-Gaussian(LG) beam can carry well defined orbital angular momentum(OAM) of light was published. An optical OAM results phase singularity of the optical field whereas the optical spin angular momentum(SAM) is related to the polarization of the field\cite{Beth1936}. The OAM of light is also termed as topological charge (TC) of optical vortex or helical phase front which can be created by dislocations in a diffractive structure. Due to its unbounded quantum numbers representing the states of OAM the LG beams have got enormous usages in fundamental studies of quantum systems \cite{MairA2001,Barriero2005,Inoue2006,Kawase2008,Dada2011,Fickler2012}, optical communications \cite{Jwang2012,willner2015,zhang2016}, detection of spinning object \cite{Lavery2013}, generation of second and higher harmonics \cite{Gariepy2014,Zhang2015,Courtial1997,Steinlechner2016,Buono2014}, wave-mixing \cite{Persuy2015}, generation of singular optical lattice \cite{Willamys2015} etc. An LG beam is employed to generate optical tweezers that can trap and rotate micron sized particles in the dark central region of the beam\cite{Friese1996,He1995,simpson1996,Gahagan1998}. This occurs due to the transfer of OAM from light to matter.
 Similar transfer has been used to create and manipulate vortex states of atomic Bose-Einstein condensate (BEC) \cite{Marzlin1997,Andersen2006,Brachmann2011,Tabosa1999,Carter2005,Alexandrescu2006,SDavis2013,mondal2014}. Under paraxial approximation OAM and SAM are decoupled. Recent works suggest that optical OAM can influence to internal electronic transitions of atoms \cite{mondal2014} only beyond the dipole transition. It is also shown that the OAM of light can be shared with quantized motion of the center of mass(CM) of atoms or molecules\cite{mondal2014}. Recent experiments on single trapped ion \cite{Ferdinand2016, Giammanco2016} have clearly demonstrated the transfer of light OAM into the internal states of the ion via quadrapole transition.  \\
In cases of trapped atoms interacting with the LG beam, the ground and first excited states of the atoms are usually involved in optical transition. The dimensions of the trap and the CM wave-functions of the atom are small compared to the transverse extent of the LG beam. As a result, the Gaussian factor of the LG field (discussed in section II), which peaks up  away  from the phase singularity of the beam(beam axis here), has always been neglected.
%It is the CM wave function not the electronic wave function which  can sense variation of electric field of the LG  beam. 
However, the change in gradient of the electric field of the beam due to presence of this Gaussian factor can generate different textures during the interaction with atoms provided the span of the atomic wave function is comparable to the beam size. Owing to the extra-ordinary width of the wave functions, Rydberg atoms seem to be well suited exploring the effects of such textures on OAM transfer in light matter interaction. 
Futher, the Rydberg atoms themselves have some special properties \cite{Gallagher1994,Robertlow2012}, such as large transition dipole moment and long life times which make them good candidates for quantum information processing \cite{Saffman2010}. Multi-wave mixing through the highly excited states of trapped Rydberg atoms is important for efficient-electromagnetically induced transparency \cite{Zhang2015a}. Population dynamics in weakly excited clouds of cold Rydberg atoms has attracted a lot recent research interests\cite{day2008, mack2015}. The interaction of the LG beam with trapped Rydberg atoms will therefore be of immense importance.\\
In the present work, we  study the interaction of LG beam with a Rydberg atom  trapped in two dimensional harmonic potential. Here we mainly analyze  the first order effect of the Gaussian factor of the beam on interaction and OAM transfer.  We consider Rydberg atoms whose radius is one order less than that of the LG beam. The  characteristic length scale of the two dimensional harmonic trap is comparable to beam waist, i.e., the characteristic length scale of the LG modes.  Here the interaction is considered on-axis. We find that the TC of the beam can be transferred directly to the Rydberg electronic state in dipole transition. As a result,  weak  electronic transition channels of atoms, like $ ^2S_{1/2} \rightarrow ^2D_{j}$, are accessible at dipole level with comparable strength as $ ^2S_{1/2} \rightarrow ^2P_{j}$ transition. This will lead to generate a final electronic state with mixed parity. Variation of strength of these transitions (in terms of the Rabi frequencies) with respect to the topological charge of the beam constitutes the central results of this work.\\
This paper is organized in following way. In section II of the paper, we develop  the theory of the transfer mechanism of the OAM in
the interaction of the LG beam with the Rydberg atoms and calculate the matrix elements of dipole transitions. We analyze our calculated Rabi frequencies  in Sec. III  with a discussion on the  numerical methods adopted to calculate the electronic and the CM wave functions. Finally, we conclude in Sec. IV with  remarks on the prospect of this work.

\section{THEORY}
The expression of electric field $ \textbf{E}(\textbf{r},t) =\mathcal{E}(\textbf{r}) e^{i \omega t} {\hat \epsilon}, $ for the  LG beam, where $\omega$ is the frequency and $\hat \epsilon$ is unit polarization vector of the beam and 
\begin{eqnarray}\label{2}
 \mathcal{E}(\textbf{r}) = \mathcal{E}_0\sqrt{\frac{2}{\pi|l|!}}\left(\frac{\rho\sqrt{2}}{w_{0}}\right)^{|l|}\exp\left(-\frac{\rho^2}{w^2_{0}}\right)e^{il\phi}e^{ikz}
 \end{eqnarray}  is the amplitude profile of the LG beam  expressed  in cylindrical coordinate ($\rho,\phi, z$) with  $l$ as the OAM\cite{SBarnett2016}. Here $\omega_0$ is the radius of the beam and the parameter $\mathcal{E}_0 = \sqrt{2I / \epsilon_0 c}$ with $I$, c, $\epsilon_0$ being the intensity, velocity of light and electric permitivity of free space, respectively. In the case of on-axis interaction, we consider that the origin of the coordinate system of the interaction is at the beam axis.
Here atomic wave functions are considered in spherical coordinate system $(\text{r},\theta,\phi)$.  The beam profile, Eq.\eqref{2}, can be expressed  in this spherical co-ordinate by replacing   $\rho = r\sin \theta$, $\phi=\phi$, $z = r\cos \theta$.
  \begin{eqnarray}\label{3}
  \mathcal{E}(\textbf{r}) &=& \mathcal{E}_0\frac{1}{w^{|l|}_{0}} \sqrt{\frac{2^{\lvert l \rvert + 1 }}{\pi  \lvert l \rvert !}} ( r \sin \theta ) ^{\lvert l \rvert } e^{il\phi}e^{-\frac{r^{2} \sin ^ 2\theta }{w^2_{0}}}e^{ikr{\cos\theta}}\nonumber \\
  &=& \mathcal{E}_0 \sum\limits_{q=0}^{\infty}f(l,q)\mathcal{R}^{l}_{|l|}(r)\mathcal{R}^{q}_{q}(r)\mathcal{R}^{-q}_{q}(r)e^{ikr\cos \theta }.
\end{eqnarray} 
Where solid harmonics $\mathcal{R}^{m}_{|l|}$ has been defined in Appendix A. Equation \eqref{3} shows two important variables: Topological charge($l$) which can be controlled externally. Another  parameter is $q$,  order  of the summation series of the exponential profile of the beam. Both the parameters of the beam can be transferred to the internal and/or the CM motions of the matter. 
We have used relations of solid harmonics here as discussed in Appendix A and obtain
$$f(l,q)=\frac{4^{q}q! }{w_0^{2q+|l|} ((2q)!)^2 |2l|!}\sqrt{(2^{(3|l|+1)} |l|!)/{\pi}}   .$$  
 The interaction Hamiltonian is given in Power-Zienau-Wooley(PZW) scheme \cite{Baiker2002,Alexandrescu2006} as
 \begin{eqnarray}\label{5}
 H_{int} = - \int \mathcal{P}(\textbf{r}^{'}) .  \textbf E (\textbf{r}^{'} , t) \mathrm{d}\textbf{r}^{'}.
 \end{eqnarray}  
The polarization vector, $ \mathcal{P}$(\textbf r), in closed integral form, is defined by 
 \begin{eqnarray}\label{6}
 \mathcal{P}(\textbf{r}^{'}) = -e \textbf{r} \frac{m_{c}}{m_{t}}\int_{0}^{1}\delta( \textbf{r}^{'} - \textbf{r}_{CM} -\lambda \frac{m_{c}}{m_{t}} \textbf{r} )
 d\lambda.
 \end{eqnarray}
 Here $m_c$ and $m_t$ are the masses of core of the atom and atom as a whole. They are considered same due to smallness of mass of the electron. The relative coordinate (internal coordinate) of the electron in the CM frame of atom is $\textbf{r} = \textbf{r}_{e} - \textbf{r}_{CM}$. 
 
 As discussed in Appendix B, the interaction Hamiltonian can be reduced to
\begin{eqnarray}\label{8c}
 H_{int} &=&  e\mathcal{E}_0 \int_{0}^{1} \sum\limits_{q = 0}^{\infty}f(l,q) A^{(l)}A^{(q)}A^{(-q)} e^{ik\lambda  r\cos \theta_e}e^{i\omega t}\ d\lambda 
\end{eqnarray} 
where 
 \begin{eqnarray}\label{8a}
 A^{(n)}\left(\lambda,r,r_{CM}\right)&=& \sum\limits_{l_1=0}^{|n|}\sum\limits_{m_1=-l_1}^{l_1}\mathcal{R}^{m_1}_{l_1}(\lambda r)\mathcal{R}^{n-m_1}_{|l|-l_1}(r_{CM})
 \end{eqnarray}
 The last angular factor $e^{ik\lambda r\cos\theta_{e}}$ in Eq. \eqref{8c} depends only on the electronic co-ordinates as  the CM motion is considered to be confined in the plane transverse to the laser propagation direction.  This angular factor can be expanded in terms of spherical harmonics: $e^{ik\lambda r\cos\theta_{e}}=\sum\limits_{p=0}^{\infty}\sqrt{4\pi(2p+1)}j_{p}(k\lambda r)Y^{0}_{p}(\theta_{e})$ with $j_{p}(k\lambda r)$ being spherical Bessel function of order $p$. $\textbf{r}\cdot \hat \epsilon$ is substituted by $r\sqrt{\frac{4\pi}{3}}\sum\limits_{\sigma=0,1,-1}^{}\epsilon_{\sigma}Y^{\sigma}_{1}(\theta_{e},\phi_{e})$, where $\epsilon_{\pm} = (E_{x} \pm i E_{y})/ \sqrt{2}$ and $\epsilon_{0} =E_{z}$. After the integration $\int_{0}^{1}\lambda^{l_1+l_2+l_3} j_p(k\lambda r)\,d\lambda $  \cite{wolfram}, the interaction Hamiltonian takes the form:

\begin{eqnarray}\label{9}
H_{int} &=& e \mathcal{E}_0  \sum\limits_{q=0}^{\infty} \frac{(\pi)^{3/2}}{\sqrt{3}}f(l,q)\sum\limits_{p=0}^{\infty} (2)^{p/4}\sqrt{2p+1}\sum\limits_{l_2,l_3=0}^{q}\sum\limits_{l_1=0}^{|l|} \Gamma\left(\frac{\alpha}{2}\right)\nonumber\\
&\times& (kr)^{\alpha}(kr_{CM})^{\beta}\times\ 
_1F_2\left(\frac{\alpha}{2};p +\frac{3}{2},\frac{1}{2}(\alpha+2);-\frac{1}{4}(kr)^{2}\right)
\nonumber\\
&\times& \sum\limits_{m_2=-l_2}^{l_2}\sum\limits_{m_3=-l_3}^{l_3}\sum\limits_{m_1=-l_1}^{l_1} \sum\limits_{\sigma=0,\pm 1}^{}\epsilon_\sigma  \mathcal{C}_{l,q.l_1,l_2,l_3}^{m_1,m_2,m_3} \mathcal{F}_e \mathcal{F}_{CM}
\end{eqnarray}
where $\mathcal{C}_{l,q.l_1,l_2,l_3}^{m_1,m_2,m_3} =\mathcal{C}^{m_{1}}_{l_{1}} \mathcal{C}^{l-m_{1}}_{\lvert l \rvert -l_{1}}\mathcal{C}^{m_{2}}_{l_{2}} \mathcal{C}^{l-m_{2}}_{q-l_2}\mathcal{C}^{m_{3}}_{l_{3}} \mathcal{C}^{-q-m_{3}}_{q-l_2},$
%$$\mathcal{F}_e=Y^{\sigma}_{1}(\theta_{e},\phi_{e})Y^{0}_{p}(\theta_{e})Y^{m_{1}}_{l_{1}}(\theta_{e},\phi_{e})Y^{m_{2}}_{l_{2}}(\theta_{e},\phi_{e})Y^{m_{3}}_{l_{3}}(\theta_{e},\phi_{e})$$
$\mathcal{F}_e=Y^{\sigma}_{1}Y^{0}_{p}Y^{m_{1}}_{l_{1}}Y^{m_{2}}_{l_{2}}Y^{m_{3}}_{l_{3}}$ and
$\mathcal{F}_{CM}= Y^{l-m_{1}}_{\lvert l \rvert -l_{1}} Y^{q-m_{2}}_{q -l_{2}} Y^{-q-m_{3}}_{q -l_{3}}$. Here, $\mathcal{F}_e$ and  $\mathcal{F}_{CM}$ are the angular components of the interaction Hamiltonian containing spherical harmonics involving the electron and the CM angular co-ordinates, respectively.
 $_{1}F_2(a;b,c;z)$ is hypergeometric function. $\alpha$ and $\beta$ are $l_1+l_2+l_3+p+1$ and $\lvert l \rvert +2q-l_1-l_2-l_3$, respectively. Here $p=0$ corresponds to dipole approximation, when the size of atom is much smaller than the wavelength of electromagnetic radiation. Since the Rydberg atom can have large size( size of Rydberg electronic wave function) that is one order magnitude less than $w_o$, it appears  that the dipole approximation breaks down for such Rydberg states. However, dipole approximation may hold good for Rydberg atoms in case of photonization occurring in the vicinity of nucleus \cite{Andersonse2015} .    
Corresponding transition matrix element (putting $p=0$ in Eq. \eqref{9}) between initial and final composite (electronic plus CM) states will be

\begin{eqnarray}\label{10}
\mathcal{M}_{i\rightarrow f} &= & \langle \Upsilon_f | H_{int} | \Upsilon_i\rangle \nonumber\\
&= & e \mathcal{E}_0  w_{r}\sum\limits_{\sigma=0,\pm 1}^{}\sum\limits_{q=0}^{\infty}\sum\limits_{l_1 =0}^{|l|}\sum\limits_{m_1 = -l_1}^{l_1}
\sum\limits_{l_2,l_3=0}^{q}\sum\limits_{m_2=-l_2}^{l_2}\sum\limits_{m_3=-l_3}^{l_3} \epsilon_\sigma \ g(l,q)\  \Gamma(\frac{\alpha}{2}) \nonumber\\
&\times & \mathcal{C}_{l,q.l_1,l_2,l_3}^{m_1,m_2,m_3} \ \ \  \langle \psi^{f}_{CM }| \left(\frac{r_{CM}}{w_r}\right)^{\beta}|\psi^{i}_{CM}\rangle \nonumber \nonumber\\
&\times& \langle \psi^{f}_e | \left(\frac{r}{w_r}\right)^{\alpha} {_1}F_{2}\left(\frac{\alpha}{2}; \frac{3}{2},\frac{1}{2}(\alpha+2);-\frac{1}{4}(kr)^{2}\right)|\psi^{i}_e\rangle \nonumber \nonumber\\
&\times &\langle Y^{m_f}_{l_f}|\mathcal{F}_e |Y^{m_i}_{l_i}\rangle
\delta_{m_1,sign(l)l_1} \delta_{m_2,l_2}\delta_{m_3,-l_3} \delta_{M_f,l-m_1-m_2-m_3+M_i}.
\end{eqnarray}
Here $$g(l,q)= \pi \left(\frac{w_r}{w_0}\right)^{2q+|l|}\frac{4^q q!}{((2q)!)^{2}|2l|!}  \sqrt{\frac{2^{3|l|+1}|l|!}{3}}$$ with 
$w_r$ is the width of the CM wavefunction, a characteristic length of trap. The last Kronecker delta function in the above matrix element represents the angular part of the CM with $M_i$ and $M_f$ as initial and final angular momenta, respectively. It appears from the term $\mathcal{F}_e$ that $q$ and $l$ have same characteristic signature which is associated with the phase gradient of wave front.  The restrictions on the other parameters, like $m_1$, $m_2$ and $m_3$ decide the new selection rule of the electronic transitions.  Due to the appearance of both $q$ and $-q$ as rank in Eq. (5), as expected, the Gaussian term of the field does not confer  net angular momentum to the system.
 The $q$ value indicates the order of $r$ and $r_{CM}$ in the expansion of Gaussian term of electric field. For example, $q=0$ designates the conventional consideration of the LG mode, like neglecting GT of the field in the $l=1$ case. Here onward we consider $_{1}F_2(a;b,c;z)\approx 1$ as other terms  involve successive power of $r^2$.  The strengths of these transitions are determined by the power of the electronic and CM coordinates, $i.e$, $\alpha$ and $\beta$. 

\subsection{Electronic dipole transition of a trapped Rydberg atom} 

The dipole transition matrix element is
\begin{eqnarray}\label{11}
\mathcal{M}_{i\rightarrow f} &= & \langle \Upsilon_f | H_{int} | \Upsilon_i\rangle \nonumber\\
&= & e \mathcal{E}_0  \sum\limits_{\sigma=0,\pm 1}^{}\sum\limits_{q=0}^{\infty}\sum\limits_{l_1 =0}^{|l|}
\sum\limits_{l_2,l_3=0}^{q} \epsilon_\sigma \ g(l,q)\  \Gamma(\frac{\alpha}{2}) \nonumber\\
&\times & \mathcal{C}_{l,q.l_1,l_2,l_3}^{m_1,m_2,m_3} \ \ \  \langle \psi^{f}_{CM }| \left(\frac{r_{CM}}{w_r}\right)^{\beta}|\psi^{i}_{CM}\rangle  \langle \psi^{f}_e | r \left(\frac{r}{w_r}\right)^{\alpha -1}|\psi^{i}_e\rangle \nonumber \nonumber\\
&\times &\langle Y^{m_f}_{l_f}|\mathcal{F}_e |Y^{m_i}_{l_i}\rangle
\ \ \delta_{M_f,l-sign(l)l_1-(l_2-l_3)+M_i}.
\end{eqnarray}
 Since $m_f=m_i+\sigma+sign(l)l_1+sign(q)(l_2-l_3)$  and minumum value of $l_f$ is $m_f$, electron receives momentum  from $l_1$, $l_2$ and  $ l_3$. It means, if component $m_1$ ($=sign(l)l_1$) of OAM goes to the internal motion, rest will  go to the CM. Whereas, if the internal motion gains $m_i(i=2,3)$ ($m_2=sign(q) l_2 $ and $m_3=-sign(q)l_3 $) from Gaussian term of the field, the CM gains $-m_i$ units angular momentum. The angular part of the electronic matrix element in Eq.(9) indicates that an arbitrary amount of the optical OAM can be transferred to the electron for various values of $m_1$ directly  without any transfer to the CM motion. Note that the typical value of $\frac{w_r}{w_o}$ is one order less than unity. Both the radial matrix elements of the electron and the CM depend on the power $\alpha$ and $\beta$, respectively, which are independent of orientation of the OAM of the beam.

 \section{NUMERICAL RESULTS AND INTERPRETATION}  
Here the CM wave function, obtained  by solving Schr\"odinger equation, of the Rydberg atom in a two dimensional harmonic  oscillatory potential is given by:
\begin{eqnarray}\label{13}
\psi_{CM}(r_{CM} , \phi)=\mathcal{A}_{N,M}(r_{CM}) e^{i M \phi} 
\end{eqnarray}
Here the normalized amplitude
   $\mathcal{A}_{N,M}(x) = \frac{1}{w_r}\sqrt{\frac{2 n_ -!}{n_+ !}} x^M L^M_{n_-}(x^2)e^{-x^2/2}$ is expressed  in terms of the characteristic coordinate ( $x=\frac{r_{CM}}{w_r}$) and
$n_{\pm} = \frac{N\pm |M|}{2}$. N is the vibrational quantum number associated with energy  $E_{CM}=
(N+1)/(w^2_r m_t)$. $M$ is the  angular momentum quantum number.  
The radial wave function of the valance electron with reduced mass $\mu$ is given by \cite{Marinescu1994}

\begin{eqnarray}\label{14}
\bigg[-\frac{\hbar^2}{2\mu}(\frac{\mathrm{d}^2}{\mathrm{d}r^2} +\frac{2}{r}\frac{\mathrm{d}}{\mathrm{d}r})+\frac{\hbar^2l(l+1)}{2\mu r^2} + V(r) \bigg] \psi_e(r) = E\psi_e(r)
\end{eqnarray}  
The potential $V(r)$ here is a sum of three physical contributions: $V(r)=V_c(r)+ V_{pole}(r)+ V_{so}(r)$, where $V_c= -\frac{Z_{nl}(r)}{r}$, with $z_{nl}$ being the effective charge from the core electron and given by $ z_{nl}(r) = 1+ (Z-1)e^{-a_1 r} -r(a_3+a_4 r)e^{-a_2 r}$ \cite{pawlak2014,Marinescu1994}.  $V_{pole}= -\frac{\alpha_c}{2r^4}(1-e^{-(\frac{r}{r_c})^6})$ is the potential due to core polarization on the valence (Rydberg) electron with $\alpha_c$ being amplitude of core polarisablity. The value of the parameters $a_1,a_2,a_3,a_4,r_c$ and $\alpha_c$ can be found in standard literature\cite{Marinescu1994}. Spin-orbit potential has well known form $V_{so}(r)=\frac{\alpha^2}{2r^3}\textbf{L}\cdot\textbf{S}$. $\alpha$ is the fine structure constant and $\langle\textbf{L}.\textbf{S}\rangle= \frac{j(j+1)-l(l+1)-s(s+1)}{2}$ with $j$, $l$ and $s$ being total angular momentum, orbital angular momentum and spin angular momentum quantum number of electron, respectively. Eq. \eqref{14}  has been solved by Numerov algorithm \cite{zimmerman1979} to obtain the radial wave function. This numerical approach requires energy values of the orbitals as input. The energy values have been calculated using the quantum defect theory \cite{Huang2010}.
 
%In calculation of transition matrix element electronic and c.m wave function have been scaled by spreading of c.m wave function.
For the numerical illustration of our work, we choose realistic parameters following the experiment of \cite{Ferdinand2016}.  
In this work, we have considered the LG beam with waist $2.7 \mu m$   
 \cite{Ferdinand2016} and intensity of $2400 V/m$ which is below ionization limit \cite{Comparat2010} of Rubidium Rydberg atom. The atom is trapped under two dimensional harmonic potential with characteristic length $w_r = 2.2 \mu m$\cite{Ferdinand2016} in the state $n^2S_{1/2,-1/2}$ (n=60). Further, first two terms (corresponding to $q$= 0 and 1) have been kept in the expansion of Gaussian factor (see Eq. \eqref{A3}) of a left circularly polarized (means $\sigma = +1$) LG beam to explore radial gradient of the field. 
Equations(5) and (6) state that beam parameters, the OAM and the order of the gradient of field (i.e. $q$ and $-q$), are shared between the motion of the electron and the CM. 
Let us first consider the case where  Gaussian factor of the beam is neglected (i.e. $q=0$) and study the generation of various channels of electronic transition as shown in TABLE 1. Simple example is the  OAM of beam be one unit ($l=1$). Following two cases arise for left and right circularly polarized light. \\
i) In the case of beam with left circular polarization, i.e. $\sigma=+1$, two possibilities of  sharing of the OAM of light can take place. First, $m_1=0$, the motion of electron does not get any contribution from optical vortex. Since   spin $m_s$ of electron can not be modified by electric field, the change of z-component of total angular momentum, $ \Delta m_{j} = m_1 + \sigma $, leads to  $S_{1/2,-1/2}\rightarrow P_{3/2,1/2}$ transition and the CM of atom gains one unit angular momentum ($M_f=1$). Here we need to keep in mind that the sign of $m_1$ will have sign of the OAM of light. The latter case, i.e., $m_1 = 1$, corresponds to transition $S_{1/2,-1/2}\rightarrow D_{5/2,3/2}$ with no angular momentum getting transferred to the CM ($M_f = 0$).\\
ii) For right circularly polarized beam, i.e. $\sigma=-1$ (see TABLE 1),  electronic transitions $S_{1/2,-1/2}\rightarrow D_{5/2,-3/2}$ and $ D_{5/2,-1/2}$ are possible corresponding  to $m_1$= 0 and 1, respectively. These transitions lead -1 and 0, respectively, as the values of OAM transferred to the CM. \\
Similarly, for negative value of the OAM of light, the CM motion of the atom acquires +1 or 0 or -1 for particular combination of polarization and $m_1$ as shown in TABLE-1.  Therefore, different electronic transition channels have emerged with different angular momentum of the CM.

Let us discuss now the situation  $q=1$ which is one of the main objectives of this work. Also, Lets us focus here only one of the cases, say, $l=+1$ and $\sigma=+1$ and this is the simplest case for $q\ne 0$.  Compare to the case of $q=0$, here we get many more channels of electronic transitions which can be understood from  Eq.\eqref{11} with different combinations of $l_2$ and $l_3$ associated with the Gaussian factor.  Though the matrix element defined in Eq. \eqref{11} symmetric with respect to $l_2$ and $l_3$, the sharing of angular momentum between the internal motion and the CM motion will be different for interchanged values of the $l_2$ and $l_3$.
Let us consider an example with $(l_1,l_2,l_3)$ as (0,1,0) and (0,0,1). Here, the OAM of light goes only to the motion of the CM as $l_1=0$.  
In the former combination,  electron will gain one positive unit from the gradient of the exponential factor of the beam. Therefore, the CM acquires one negative unit from the gradient which nullifies the OAM of the CM obtained from the OAM of light.  In the latter combination, the CM acquires positive unit of OAM (i.e., the electron gains one negative unit from the Gaussian factor of the beam) to have total OAM the CM equals to 2. But, in both the cases transitions will be  $S_{1/2,-1/2} \rightarrow D_{5/2, -1/2+\Delta m_j} $ transition. Now, the change $\Delta m_j$ will be  decided from the value of $m_2+m_3+\sigma$, which generate $\Delta m_j=2$  and $0$ 
in the above cases. respectively. 
It is clear from the TABLE II, Rabi frequency in the $S\rightarrow D$ transition corresponding to $\Delta m_j = 2 $ is greater than  that of $ \Delta m_j = 0$. Therefore, we can predict that the Rabi frequencies for $\Delta m_j = -2$ will be greater than that of $\Delta m_j = 0$ when both the $(l,\sigma)$  angular momentum are  negative. 

Similarly, channels $\Delta m_j = 0$  can be explored by making opposite signs of $\sigma$ and $l$. So, depending on the relative sign of $\sigma $ and $l$, most of the magnetic sub-levels in $S\rightarrow D$ transitions\cite{Ferdinand2016},  can be explored by LG beam in dipole level. It is possible to get $\Delta m_j=1$ by either choosing linear polarization or transfer of OAM of light totally in the motion of the CM.

\begin{table}
\caption{Various possible final states depending on sign of topological charge and polarization neglecting Gaussian contribution }
\centering
\centering
\begin{tabular}{c c c c c c c  }
$l $ & $\sigma $ & $ m_1 $ & $M_f$& Initial state & Final state &\\
\hline
1 & 1 & 0 & 1 & $ S_{1/2,-1/2}$ & $P_{3/2,1/2}$ & \\
   &    & 1 & 0 & $ S_{1/2,-1/2}$ & $D_{5/2,3/2}$ & \\
1 & -1& 0 & -1  &$ S_{1/2,-1/2}$ & $D_{5/2,-3/2}$ &\\
   &    & 1 & 0 &$ S_{1/2,-1/2}$ & $D_{5/2,-1/2}$ &\\
-1 & 1 & 0 & 1 &$ S_{1/2,-1/2}$ & $D_{5/2,1/2}$ &\\
    &    &  -1 & 0 &$ S_{1/2,-1/2}$ & $D_{5/2,-1/2}$ &\\
-1 & -1 & 0 & -1 &$ S_{1/2,-1/2}$ & $P_{3/2,-3/2}$ & \\
    &    & -1 & 0 & $ S_{1/2,-1/2}$ & $D_{5/2,-5/2}$ &\\ 
\end{tabular} 

\end{table}

\begin{figure}[h]
\centering
\includegraphics[width=1.2\textwidth]{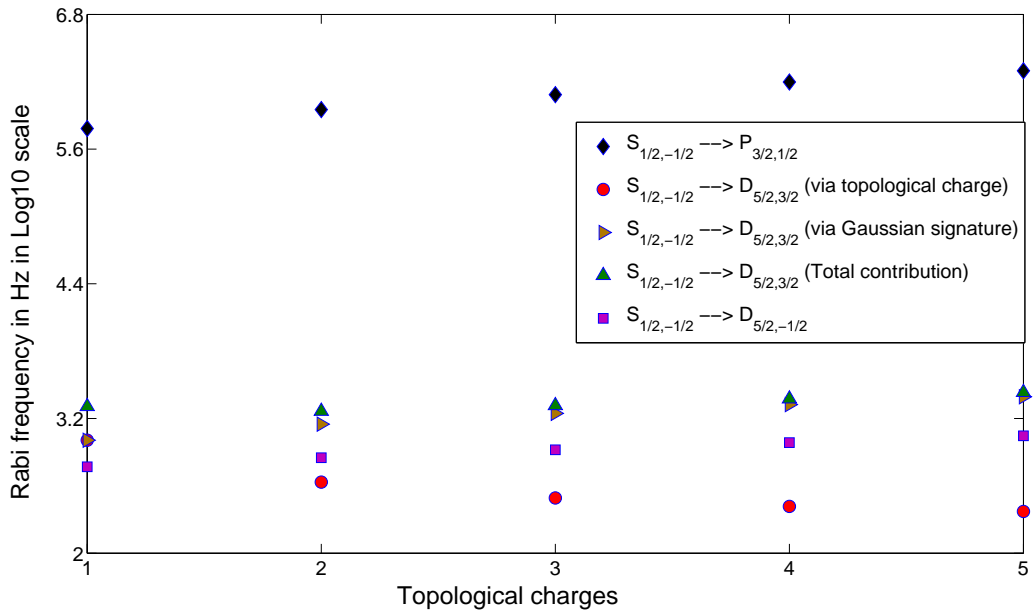}
\caption{(color online) Variation of Rabi frequency with topological charge for various transitions.  }
\label{fig:2}
\end{figure}

  Figure \ref{fig:2} shows the behavior of the Rabi frequency for different transition channels with respect to topological charge of the beam. Interestingly, significant increase of the Rabi frequency is seen for  $S\rightarrow P$ channel with the increase of the charge, however the frequency decreases  for the $S\rightarrow D $ channels.  This will be cleared if we look into the plot and compare channel $S\rightarrow P$ with $S\rightarrow D_{5/2,3/2}$ ("via TC"), i.e., when $l_1=m_1=1$ and all other $l_i=m_i=0$. Physically, latter channel represents the gain of the angular momentum  by the electron from light beam, which is not true for the former channel.
All the values of radial matrix elements along with the angular coefficients, shown in Eq.(9),  increase with the charge of optical vortex  apart from the factor $g(l,q)$. Therefore, it is the competition between $g(l,q)$ and  the rest of the terms that decide the overall trend of the Rabi frequency. As shown in TABLE II, transition channel $S\rightarrow D_{5/2,3/2}$ will also be produced from $l_1=0$, but with $l_2=1, l_3=0$. This means that electron will not gain angular momentum from the topological charge of beam, but from the gradient of the Gaussian profile (i.e. $q\neq 0$) of the beam (it is designated as   'via GT' in the plot). Unlike 'via TC', the Rabi frequency for 'via GT' channel will increase with $l$ value, which makes the total Rabi frequency of $S\rightarrow D_{5/2,3/2}$(total) channel is more or less constant. There will be other possibilities of internal transitions here with larger change of orbital angular number, but their effects are negligible due to very weak Rabi frequencies. \\

\begin{table}
\caption{Dipole Rabi frequencies for various electronic channels depending on OAM transferred to electron.}
\centering
\centering
\begin{tabular}{c c c c c c c c c c }
 $ l_1 $ & $ l_2 $ & $ l_3 $ & $m_1$ & $m_2$  & $m_3$ & $M_{f}$ & transition channels & Rabi frequency(in KHz ) \\
\hline
 0 & 0 & 0 & 0 & 0 & 0 & 1 & $S_{1/2,-1/2}\rightarrow P_{3/2,1/2}$ & $607$ & \\
   1 & 0 & 0 & 1 &  0 & 0 &  0& $S_{1/2,-1/2}\rightarrow D_{5/2,3/2}$ &$1.01$ & \\
   
    0 & 1 & 0 & 0 &  1 & 0 & 0 &$S_{1/2,-1/2}\rightarrow D_{5/2,3/2}$ & $1.01$ &
             \\
     0 & 0 & 1 & 0 &  0  & -1  &  2 & $S_{1/2,-1/2}\rightarrow D_{5/2,-1/2}$ &  $0.59$ & \\
\hline
\end{tabular}
\end{table}
Similarly, we can shine beam with both the OAM and SAM of the field having negative unity values. But, interesting physics can be observed when the OAM and the SAM have opposite sign with  $l_1=0 $ and $q\neq 0$. Table III shows a unique situation where by changing relative orientation of the OAM and SAM of the beam, we can tune particular internal transition along with distinct angular momentum of the CM motion. This can be experimentally verified by the procedure discussed in Reference \cite{Schmiegelow2012}.
Also, it is possible to create entanglement between internal and external angular momentum in the final states \cite{Muthukrishnan2002}.\\

\begin{table}
\caption{Dipole Rabi frequencies for various electronic channels depending on OAM transferred to electron.}
\centering
\centering
\begin{tabular}{c  c  c c c c c c c c c c }
$ l$ & $\sigma$ & $ l_1 $ & $ l_2 $ & $ l_3 $ & $m_1$ & $m_2$  & $m_3$ & $M_{f}$ & transition channels & Rabi frequency(in KHz) \\
\hline
-1 & 1 &   0 & 1 & 0 & 0 &  1 & 0 & -2 &$S_{1/2,-1/2}\rightarrow D_{5/2,3/2}$ & $1.01$ &
             \\
1  & -1 & 0 & 0 &  1 & 0 & 0 &  -1  &  2 & $S_{1/2,-1/2}\rightarrow D_{5/2,-3/2}$ &  $1.01$ & \\
\hline
\end{tabular}
\end{table}

\section{CONCLUSION}
Here we  have studied how spatial structure of vortex light  couples to both internal electronic and external CM motion of trapped Rydberg atoms. We have treated the electronic and the CM coordinate on equal footings. The OAM of field  can be transferred directly to the electronic motion which modifies dipole selection rule opening up a plethora of transition channels with different Rabi frequencies. Distinct identification of these channels may be possible either by external magnetic field , or by LG beam induced magnetism \cite{Quinteiro}. Our analysis shows that the large  size of a Rydberg atom and the extended CM wave-function lead to an appreciable effect of the Gaussian factor of the beam in the light matter interaction. This part of interaction will be significant and at par   
with the interaction arising from the inherent vorticity of optical field. As the field OAM is shared between the electron and the CM, an entanglement induced by the LG beam is  inevitable between the combined final states of the electron and the motion of the CM at the level of  $S\rightarrow D$ transition. Another salient feature of this light matter interaction is generation of mixed parity state, i.e. $\psi = \alpha|P_{3/2,1/2}> + \beta |D_{5/2,3/2}> + \gamma|D_{5/2,-1/2}>+...  $. These mixing coefficients depend on the Rabi frequencies which depend on the vortex charge of LG beam. Further, we show here the possibility of tuning or controlling different channels of internal transitions by proper choice of the OAM and the SAM of the field.  It would be interesting to have similar study in future when the LG beam is focused \cite{anal2016}.\\
Acknowledgments: we are thankful to Anal Bhowmik for fruitful discussions and comments on the manuscript.

\appendix*
\section{A: Relations of Solid Harmonics}

 The solid harmonics $\mathcal{R}^{l}_{|l|}$ is  defined by 
\begin{equation}\label{A1}
\mathcal{R}^{m}_{l} = \mathcal{C}^{m}_{l}r^{l}Y^{m}_{l}(\theta,\phi)
\end{equation}
 with 
 $\mathcal{C}^{m}_{l}=\sqrt{[4\pi/(2|l|+1)/(l-m)!/(l+m)!]}$.
 
We can express the following factors in terms of solid harmonic  \cite{van1998}.
\begin{equation}\label{A2}
(r\sin \theta)^ {\lvert l \rvert } e^{i(\pm l)\phi } = (\pm)^l 2^{\lvert l \rvert } \frac{\lvert l \rvert !}{\lvert 2l \rvert !} \mathcal{R}^{\pm l}_{\lvert l \rvert}(r)
\end{equation}
 and
\begin{equation}\label{A3} 
 e^{-r^{2}\sin^2\theta/w^2_{0} }=\sum\limits_{q=0}^{\infty}\frac{1}{w^{2q}_0}4^{q}\frac{q!}{(2q!)^2}\\ 
\mathcal{R}^{q}_{q}(r)\mathcal{R}^{-q}_{q}(r).
 \end{equation}

\section{B}
  
 Substituting Eq. \eqref{3} and Eq. \eqref{6} into \eqref{5}, $H_{int}$ takes the form 
\begin{eqnarray}\label{7}
 H_{int} = e\frac{m_{c}}{m_{t}}\textbf{r}.\epsilon \int\limits_{0}^{1}\mathcal{E}(\textbf{r}_{c.m}+\lambda\frac{m_{c}}{m_{t}}\textbf{r})e^{i\omega t}\,d\lambda
 \end{eqnarray}
 
 Eq. (3) and Eq. (6) show that now we require solid harmonics of the form
 $\mathcal{R}^{m}_{l}(r_{c.m}+\lambda\frac{m_{c}}{m_{t}}r)$.  
 Additional theorems of regular solid harmonics  can be used to separate internal and external coordinates as 
\begin{eqnarray}
 \mathcal{R}^{m}_{l}(r_{c.m}+\lambda\frac{m_{c}}{m_{t}}r)& = & \sum\limits_{l_{1}=0}^{|l|}\sum\limits_{m_{1}=-l_{1}}^{l_{1}}\mathcal{R}^{m_{1}}_{l_{1}}(\lambda\frac{m_{c}}{m_{t}}r)\mathcal{R}^{l-m_{1}}_{|l|-l_{1}}(r_{c.m}) \nonumber \\
 &=&\sum\limits_{l_{1}=0}^{|l|}\sum\limits_{m_{1}=-l_{1}}^{l_{1}}C^{m_{1}}_{l_{1}}(\lambda\frac{m_{c}}{m_{t}}r)^{l_{1}}
 Y^{m_{1}}_{l_{1}}(\theta_{e},\phi_{e})\mathcal{C}^{l-m_{1}}_{|l|-l_{1}} r_{c.m.}^{|l|-l_{1}}Y^{l-l_{1}}_{|l|-l_{1}}(\theta_{c.m},\phi_{c.m.}).
\end{eqnarray} 

%\section{ACKNOWLEDGMENT}

%) ============================================================================
% === REFERENCES =============================================================
% ============================================================================

%\bibliographystyle{abbrv}

%\bibliography{bib}

\end{document}